\documentclass{elsart}
\usepackage[]{latexsym,amsmath,amssymb}
\usepackage[]{epsfig}
\newcommand{\be}{\begin{eqnarray}}
\newcommand{\ee}{\end{eqnarray}}
\newcommand{\bra}[1]{\mbox{$\langle\, #1 \mid$}}

\newcommand{\ket}[1]{\mbox{$\mid #1\,\rangle$}}

\newcommand{\pro}[2]{\mbox{$\langle\, #1 \mid #2\,\rangle$}}
\newcommand{\expec}[1]{\mbox{$\langle\, #1\,\rangle$}}

\begin{document}
\begin{frontmatter}
\title{\bf On the gravitational collapse in anti-de~Sitter
space-time}
\author[a]{G.L.~Alberghi\thanksref{aa}}
\author[a]{R.~Casadio\thanksref{bb}}
\address[a]{Dipartimento di Fisica, Universit\`a di Bologna, and
I.N.F.N, Sezione di Bologna,
Via Irnerio 46, 40126 Bologna, Italy.}
\thanks[aa]{E-mail: alberghi@bo.infn.it}
\thanks[bb]{E-mail: casadio@bo.infn.it}
\begin{abstract}
We study the semiclassical evolution of a self-gravitating thick
shell in Anti-de~Sitter space-time.
We treat the matter on the shell as made of quantized bosons and
evaluate the back-reaction of the loss of gravitational energy
which is radiated away as a non-adiabatic effect.
A peculiar feature of anti-de~Sitter is that such an emission also
occurs for large shell radius, contrary to the asymptotically flat
case.
\end{abstract}
\begin{keyword}
Radiating this shell \sep Anti-de~Sitter space-time \sep
Semiclassical Gravity
\PACS 04.40.-b \sep 04.70.Dy \sep 98.70.Rz
\end{keyword}
\end{frontmatter}
One of the main issues arising in quantum gravity is the search for a
unitary description of the process of black hole formation and evaporation
\cite{hawking}.
The conjectured AdS-CFT correspondence may lead to a way of solving
this puzzle, as it provides a gauge theory (hence unitary) description
for processes occurring in string theory and asymptotically
anti-de~Sitter (AdS) space-times \cite{maldacena}.
So far, however, it has been difficult to find a concrete description
of such a process within this framework.
For this reason, one is led to examine simplified models, for black
hole formation, such as the collapse of a spherical shell of matter,
in order to get an insight of what a full quantum gravity
description would be.
This is the path followed in Refs.~\cite{giddings,Trodden,vakkuri}.
The main unsolved issue in these analysis is the description
of how the initial stages of the evolution, for a shell with a
radius much greater then its horizon, may be related to the
late stages of the collapse, when a black hole is about to be
formed.
In Ref.~\cite{giddings}, it is suggested that a solution may be
provided by taking into account the radiation emitted by the shell
as it collapses.
This is just what we are going to describe in the following.
\par
In a series of papers \cite{acvv1,acvv2}, we have studied
the semiclassical dynamics of a self-gravitating shell
made of bosonic matter (modelled as a set of $N\gg 1$
``microshells'') which collapses in an asymptotically flat
(Schwarz\-schild) space-time.
The (average) radius of the shell is taken to be a function
of (proper) time $r=R(\tau)$, and its thickness $\delta$ is small
but finite ($0<\delta\ll R$) throughout the evolution.
The main result of Ref.~\cite{acvv2} was that tidal forces
acting inside the shell induce non-adiabatic (i.e.,
proportional to the contraction velocity) changes in the
quantum mechanical states of the microshells and give rise
to a probability of excitation which becomes appreciable as
the shell approaches its own gravitational radius
$R_H=2\,M_s$, where $M_s$ is the Arnowitt-Deser-Misner
(ADM) mass of the shell~\footnote{With use units with $c=1$
and $\ell_p^2=\hbar\,G_{\rm N}$ is the Planck length squared.}.
If the bosonic microshells are coupled to a radiation field,
they will then emit the excess energy and the ADM mass $M_s$
will steadily decrease.
This induces an effective (quantum) tension which will modify
the trajectory of the shell radius.
We explicitly determined the evolution as long as $R$
is not too close to $R_H$, where some of the approximations
employed break down, and showed that the shell contracts
at constant (terminal) velocity.
\par
In the present paper we want to generalize our previous results
considering a shell embedded in asymptotically AdS space-time
with cosmological constant $\Lambda=3/\ell^2$ (for the thermodynamics
of such a case see Ref.~\cite{acv2}).
As we shall show below, one expects strong non-adiabatic
effects not just at small shell radius (when $R\sim 2\,M_s$ and the
self-gravity of the shell dominates), but also for $R\gg \ell$
(when it is the cosmological constant that mainly determines the
shell motion).
In particular, we will focus on $R$ large, in order to highlight
the differences with respect to the case already analyzed
in Refs.~\cite{acvv1,acvv2}.
\par
The Schwarzschild-AdS metrics inside and outside the shell at
$r=R(\tau)$ can be written as
\be
ds^2=-f_{\rm i/o}\,dt^2
+f^{-1}_{\rm i/o}\,dr^2
+r^2\,d\Omega^2
\ ,
\ee
where $d\Omega^2=d\theta^2+\sin^2\theta\,d\phi^2$,
$\tau$ is the shell proper time,
\be
d\tau\equiv \sqrt{f_{\rm o}(R)}\,dt
=\sqrt{f_{\rm i}(R)}\,dt
\ ,
\label{dt}
\ee
and
\be
f_{\rm i/o}(r)=1-{2\,M_{\rm i/o}\over r}+{r^2\over\ell^2}
\ .
\ee
with $M_{\rm i/o}$ the ADM masses for $r<R$ and $r>R$
respectively.
A straightforward application of the junction conditions \cite{israel}
yields the equation of motion for the shell,
\be
\dot R^2=\left({M_{\rm o}-M_{\rm i}\over M}\right)^2
+{M_{\rm o}+M_{\rm i}\over R}
+{M^2\over 4\,R^2}
-{R^2\over \ell^2}
-1
\ ,
\label{Req}
\ee
where a dot denotes derivative with respect to $\tau$.
The proper mass $M$, as well as $M_{\rm i/o}$, may depend on $\tau$
\cite{acv}.
In particular, we shall consider $M$ constant~\footnote{As in
Ref.~\cite{acvv2}, we are interested in the loss of gravitational energy
from the shell.
Allowing for a decrease of the proper mass $M$ would further strengthen
the effect, and will be the subject of future works.}
and initial conditions at $\tau=0$ such that $\dot R(0)=0$ [$R(0)$
equals the classical turning point].
In order to compare with the asymptotically flat case (obtained as
$\ell\to\infty$), we shall also set $M_{\rm i}=0$ and consider the
range of parameters
\be
\delta\ll M\lesssim M_s(0)\ll \ell \ll R(0)
\ ,
\label{range}
\ee
where $M_s\equiv M_{\rm o}(\tau)$.
Note also that $R(0)\gg\ell$ requires $M_s(0)/M\sim R(0)/\ell$.
\par
Let us now note that the proper acceleration of the shell radius
given by
\be
\ddot R=-{M_s\over 2\,R^2}-{M^2\over 4\,R^3}-{R\over\ell^2}
\ ,
\label{ddR}
\ee
is dominated by the last term for large $R$.
One then obtains that, for $R\gg\ell$ and near the turning point,
the acceleration of the shell is much larger than it
would be in asymptotically flat space ($\ell\to\infty$), so that
the contraction velocity increases faster and non-adiabatic
effects might occur before $R$ approaches $R_H$.
\par
The motion of a microshell in the system of $N\gg 1$ microshells
can be determined from the junction equation (\ref{Req}).
In particular, one can obtain a mean field Hamiltonian equation
\cite{acvv1,acvv2} for the relative displacement $\bar r$ of
each microshell of proper mass $m$ with respect to the average
radius $R$ of the shell, which takes the form
\be
H\equiv{1\over 2}\,(M+m)\,\dot R^2+{1\over 2}\,\mu\,\dot{\bar r}^2
+V_0+V_m(\bar r)=0
\ ,
\label{H}
\ee
where $\mu\equiv m\,(M-m)/M$ is the effective mass of the
microshell and $V_0$ contains all the terms which do not depend
on $\bar r$.
Since $m\ll M=N\,m$, it is consistent just to retain terms
linear in $m$ and approximate $\mu\simeq m$.
Further, we are interested in the case when the shell radius
is large and its thickness small [see Eq.~(\ref{range})].
The time evolution of $R$ is thus determined by
\be
{1\over 2}\,M\,\dot R^2\simeq -V_0
\ ,
\label{HJ}
\ee
and that of $\bar r$ by
\be
{1\over 2}\,m\,\dot{\bar r}^2\simeq -V_m(\bar r)
\ .
\label{Hm}
\ee
The potential, to first order in $\bar r$, is given by
\be
V_m={M_s\,m\over 2\,R}\,{\bar r\over R}\times
\left\{
\begin{array}{ll}
\strut\displaystyle\left(1-{M^2\over 2\,R\,M_s}\right)
&
\ \ \ \ \ \ \
\bar r>+{\delta\over2}
\\
&
\\
\strut\displaystyle\left(-1-{M^2\over 2\,R\,M_s}\right)
&
\ \ \ \ \ \ \
\bar r>+{\delta\over2}
\ .
\end{array}
\right.
\ee
Up to corrections of order $\bar r^2/\ell^2$, which are negligible
in the range of parameters (\ref{range}), the above expression
is identical to the asymptotically flat case treated in
Ref.~\cite{acvv2} and represents the tidal force responsible for the
microshell confinement.
This is expected in a proper time formalism, since the binding
potential $V_m$ just depends on the local geometry.
\par
The Hamiltonian constraint (\ref{Hm}) can be quantized and,
via the Born-Oppenheimer approximation, on assuming negligible
quantum fluctuations of the metric, it leads to the
Hamilton-Jacobi equation (\ref{HJ}) for the average
radius $R$ and to a Schr\"odinger equation for the
microshell~\footnote{Such a treatment is rather lengthy and we
omit the details for the sake of brevity. For the general formalism
see Ref.~\cite{brout}; its application to the shell model can
be found in Ref.~\cite{acvv1,acvv2}.},
\be
i\,\hbar\,\partial_\tau\ket{n}=
\left[-{\hbar^2\over 2\,m}\,\partial_{\bar r}^2+V_m\right]\,\ket{n}
\ ,
\ee
in which the potential $V_m$ is time-dependent (recall that $R$
and $M_s$ are in general functions of $\tau$).
Such equation was analyzed in Ref.~\cite{acvv2} and
the low lying modes in the spectrum,
\be
\pro{\tau,\bar r}{n}=e^{-i\,n\,\Omega\,\tau}\,\Phi_n(\bar r)
\ ,
\ee
were obtained by means of quantum invariants \cite{lewis}.
The low energy levels are given by
\be
E_n-E_0\simeq n\,\hbar\,\Omega
\ ,
\ee
where the fundamental frequency is
\be
\Omega={1\over R}\,\sqrt{M_s\over \delta}
\ ,
\ee
and the spatial width of the levels is
\be
\delta\sim
\ell_m^{2/3}\,R^{1/3}\,\left({R\over M_s}\right)^{1/3}
\ ,
\ee
$\ell_m=\hbar/m=N\,\ell_p^2/M$ being the Compton wavelength of the
microshells.
\par
Due to the time-dependence of $V_m$, the amplitude of
excitation from the ground state $\Phi_0$ to states of even
quantum number $\Phi_{2n}$ at the time $\tau>0$ is not
zero.
To leading order in $\dot R$, it is given by
\be
A_{0\to 2n}(\tau)\simeq
(-i)^n\,{\sqrt{(2\,n)!}\over 6^n\,n!}\,
\left({\delta\over M_s}\right)^{n/2}\,\dot R^n
\ .
\ee
This expression rapidly vanishes for large $R$
($\gg M_s$) in the asymptotically flat case.
However, in AdS, $|\dot R|$ grows faster for $R\gg\ell$
[as can be inferred from Eq.~(\ref{ddR})] and leads to a non
negligible amplitude in such a regime.
\par
The microshells can be coupled to an external scalar field
$\varphi$ by an interaction Lagrangian density of the form
\be
\hat L_{\rm int}=
%-
e\!\!\!\sum_{n,n'\not=n}\!\!\!
e^{-i\,(n-n')\,\Omega\,\tau}\,
\Phi_n(\bar r)\,\Phi_{n'}(\bar r)\,\hat\varphi(t,R+\bar r)
\ ,
\ee
where $e$ is the coupling constant and $\tau=\tau(t)$ [see
Eq.~(\ref{dt})].
The field $\varphi$ satisfies the Klein-Gordon equation
$\Box\varphi=0$ everywhere in AdS.
However, since we are interested in the flux of radiation
outgoing from the shell, we just need solutions outside
the shell and neglect the time-dependence of $R$ (consistently
with the approximation in which we keep the leading term for
small shell velocity).
The scalar field for $r>R$ is given by a sum over
the normal modes \cite{giddings}
\be
\pro{t,r,\theta,\phi}{n,l,m}
={e^{-i\,\omega_{nl}\,t}\over r}\,u_{nlm}(r)\,
Y_l^m(\theta,\phi)
\ ,
\ee
where the $Y$'s are standard spherical harmonics.
The radial functions are more easily expressed in terms
of the turtle-like coordinate $dr_*\equiv dr/f_{\rm o}$, and read
\be
u_{nlm}=c_{nl}\,\cos^2(r_*)\,\sin^{1+l}(r_*)\,
P_n^{l+{1\over 2},{3\over 2}}\left(\cos(2\,r_*)\right)
\ ,
\ee
where the $c_{nl}$'s are normalization constants, the $P$'s
Jacobi polynomials and
\be
\omega_{nl}=\,\sqrt{{f_{\rm o}(R)\over 1+R^2/\ell^2}}\,
{2\,n+l+3\over \ell}
\ .
\ee
\par
We may now compute the transition amplitude for the entire
process during which a microshell is excited from the ground
state $\Phi_0$ to an excited state $\Phi_n$ and then
decays back to $\Phi_0$ by emitting scalar quanta $\ket{n,l,m}$,
\be
\!\!\!\!\!\!\!
{\mathcal M}_n(t,0)\equiv
{i\over\hbar}\,\int_{0}^{t} dt'\,\sum_{i=1}^N\,
\int d\bar r'_i\,\Phi_0^*(\bar r'_i)\,
\Phi_0^{\phantom{a}}(\bar r'_i)\,
\bra{n,0,0}\hat L_{\rm int}(t',\bar r'_i)\ket{0,0,0}\,
\ ,
\ee
where we enforced the spherical symmetry (so that only states with
$l=m=0$ are included) and summed over the $N$ microshells located
at $R+\bar r_i$.
We further assume that $A_{0\to 2}(t)$ is the only non-negligible
excitation amplitude (that is, we neglect the
possibility that each microshell can be excited more than once)
and consider a symmetrized state for the $N$ microshells.
The probability of emitting an energy equal to $2\,\Omega$ during
the interval between $t=0$ (when $\dot R=0$) and $t$,
\be
P(2\,\Omega;t)\simeq
\sum_{n'\,n''}\,
{\mathcal M}^*_{n'}(t,0)\,{\mathcal M}_{n''}^{\phantom{a}}(t,0)
\ ,
\ee
contains the projection of the Wightman function onto states with
$l=m=0$,
\be
\expec{\hat\varphi(R')\,\hat\varphi(R'')}
&=&
\sum_n\,{e^{-i\,\omega_{n0}\,(t'-t'')}\over 2\,\omega_{n0}\,R'\,R''}
\,u_{n00}^{\phantom{a}}(R_*')\,u_{n00}^*(R_*'')
\nonumber
\\
&\equiv&
\sum_n\,{e^{-i\,\omega_{n0}\,(t'-t'')}\over 2\,\omega_{n0}\,R'\,R''}
\,s_n^2(R',R'')
\ ,
\ee
and is finally given by
\be
P(2\,\Omega;t)\sim
{e^2\,N^2\over\hbar^2}\,|A_{0\to 2}|^2\,
\sum_n\,s_n^2(R,R)\,
{\sin^2\left[\left(2\,\Omega\,\sqrt{f_{\rm o}}-\omega_{n0}\right)
\,t/2\right]
\over\,\omega^2_{n0}\,
\left(2\,\Omega\,\sqrt{f_{\rm o}}-\omega_{n0}\right)^2}
\ ,
\label{P2O}
\ee
In the above expression, all functions must be evaluated at the time
$t$ and we also used the fact that the states $\Phi_n$ have a
very narrow width $\delta$ [according to Eq.~(\ref{range})]
and can be approximated as $\Phi_n\sim\delta(\bar r)$.
Note that the lapse $t$ should be chosen short enough so that
$R$ remains approximately constant during the corresponding time
interval, but long enough to average over transient
effects~\footnote{This is the approximation which yields the
Planckian spectrum at the Hawking temperature for $R\sim 2\,M_s$
\cite{acvv2}.}.
\par
\begin{figure}[t]
\centering
\raisebox{3cm}{$R$}\hspace{0cm}
\epsfxsize=3in
\epsfbox{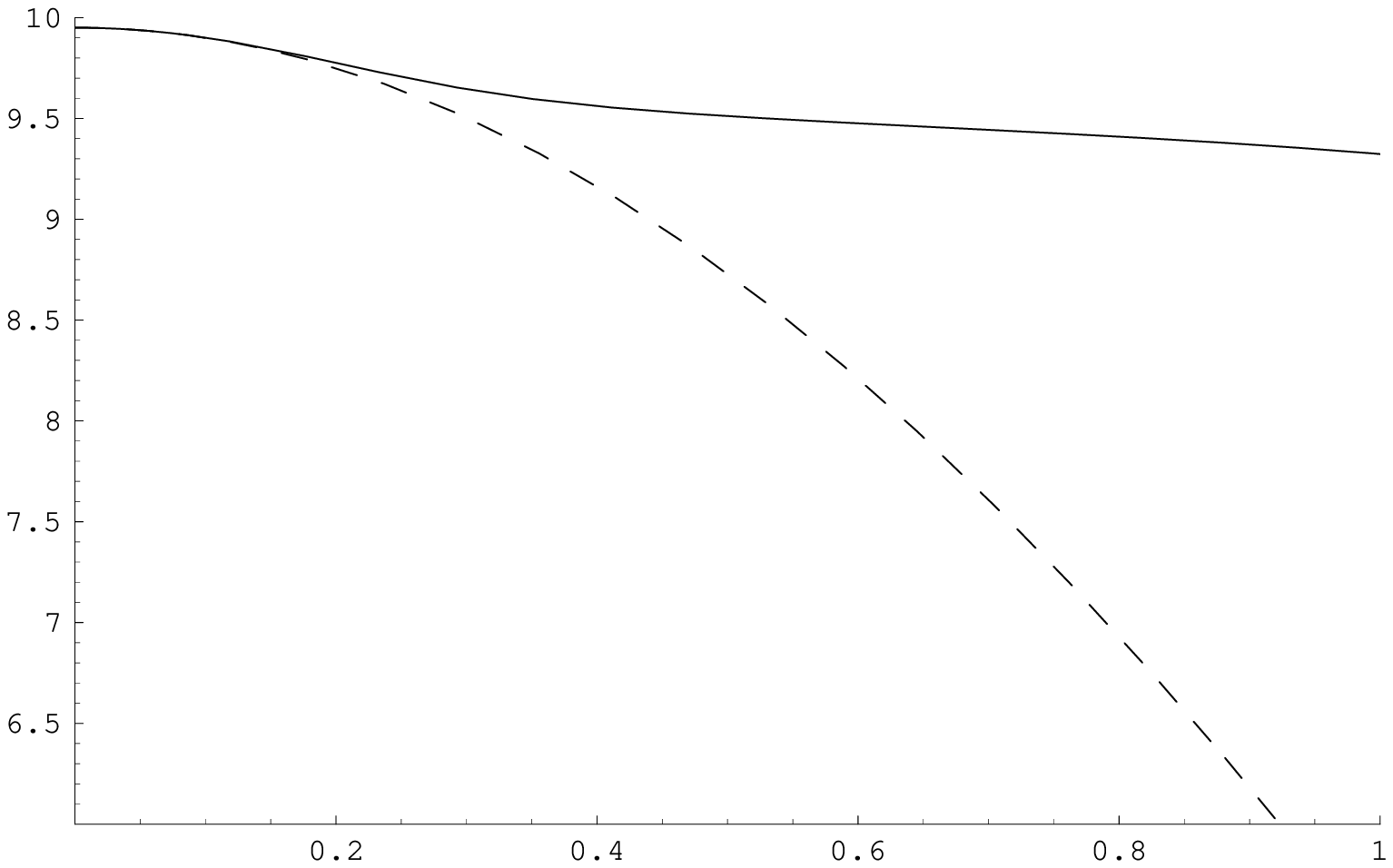}
\\
\raisebox{0cm}{\hspace{6cm} $\tau$}
\\
\raisebox{0cm}{\hspace{1cm} (a)}
\\
\ \phantom{a}
\\
\raisebox{0cm}{\hspace{6cm} $\tau$}
\\
\raisebox{3cm}{$\dot R$}\hspace{0cm}
\epsfxsize=3in
\epsfbox{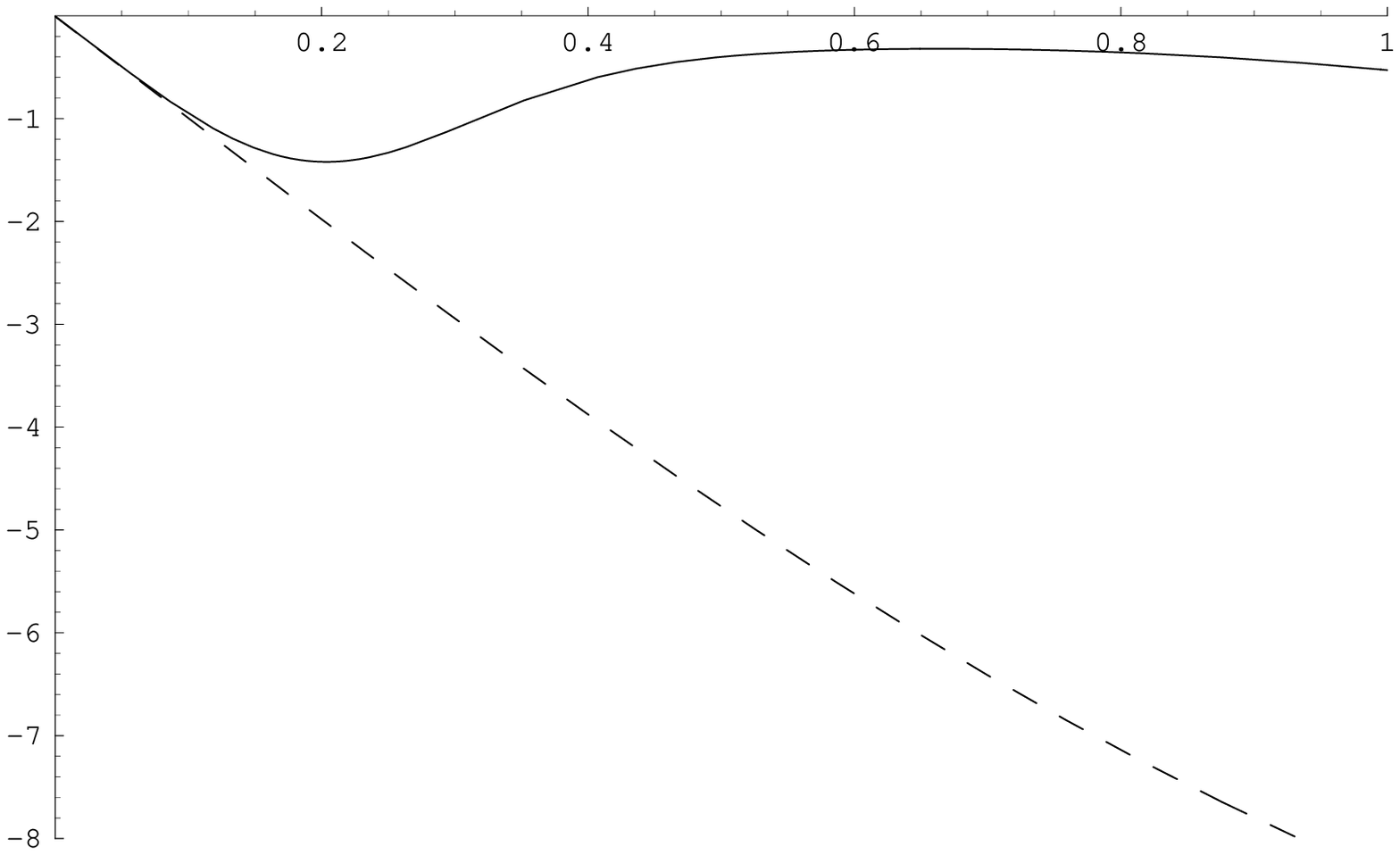}
\\
\raisebox{0cm}{\hspace{1cm} (b)}
\\
\ \phantom{a}
\\
\raisebox{0cm}{\hspace{6cm} $\tau$}
\\
\raisebox{3cm}{$\bar M_s$}\hspace{0cm}
\epsfxsize=3in
\epsfbox{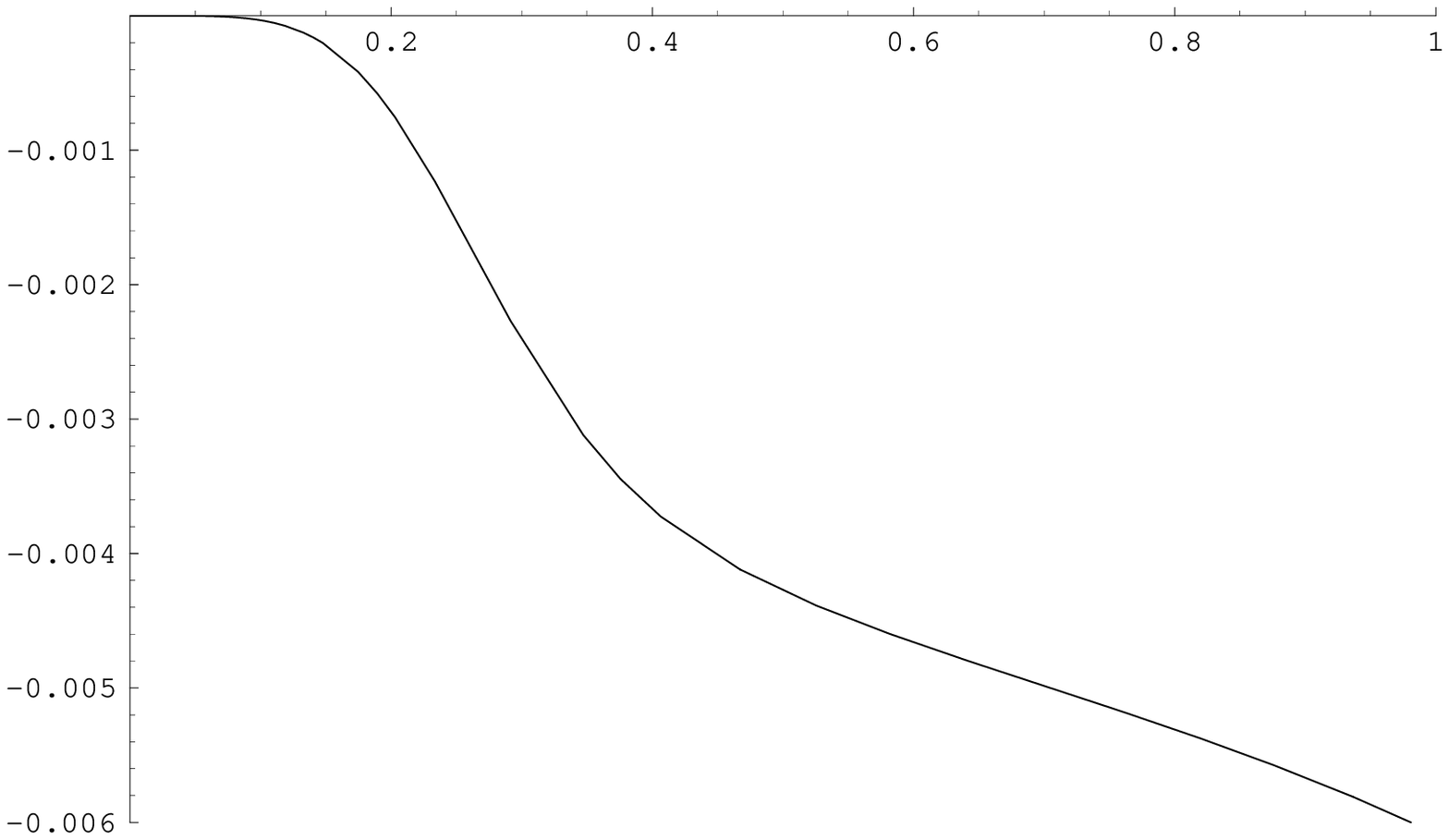}
\\
\raisebox{0cm}{\hspace{1cm} (c)}
\caption{(a) shell radius with (solid line) and without (dashed line)
radiation;
(b) shell velocity with (solid line) and without (dashed line)
radiation;
(b) shell ADM mass $\bar M_s\equiv M_s-M_s(0)$.
\label{R}}
\end{figure}
\begin{table}
\centering
%\begin{ruledtabular}
\begin{tabular}{|c|c|c|c|c|c|c|c|c|c|}
\hline
$\delta$ & $M$ & $M_s$ & $\ell$ & $R$ &
$N$ & $n$ & $\Omega$ &
$\displaystyle{|A_{0\to 2}|^2\over\dot R^2}$ &
$\displaystyle{\Delta M_s\over M_s}$
\\
\hline
$10^{23}$ &
$10^{24}$ &
$10^{25}$ &
$10^{26}$ &
$10^{27}$ &
$10^{44}$ &
$10^2$ &
$10^{-24}$ &
$10^{-2}$ &
$10^{-7}$
\\
\hline
$10^{25}$ &
$10^{26}$ &
$10^{28}$ &
$10^{30}$ &
$10^{32}$ &
$10^{46}$ &
$4\cdot 10^{2}$ &
$10^{-28}$ &
$10^{-3}$ &
$10^{-12}$
\\
\hline
$10^{26}$ &
$10^{29}$ &
$10^{31}$ &
$10^{33}$ &
$10^{35}$ &
$10^{49}$ &
$4\cdot 10^{4}$ &
$10^{-29}$ &
$10^{-5}$ &
$10^{-15}$
\\
\hline
$10^{27}$ &
$10^{32}$ &
$10^{34}$ &
$10^{36}$ &
$10^{38}$ &
$10^{52}$ &
$4\cdot 10^{6}$ &
$10^{-30}$ &
$10^{-7}$ &
$10^{-18}$
\\
\hline
\end{tabular}
%\end{ruledtabular}
\caption{The parameters $\ell$, $M$ and $N$ and relevant
quantities at $\tau=0$ for $\ell_m=10^{20}$.
The integer $n$ identifies the AdS mode corresponding to the
excited energy gap $2\Omega$ and the change $\Delta M_s$ is
estimated assuming $\dot R^2\sim 1$ and neglecting the time of
emission.
All lengths are in units of $\ell_p$.}
\label{T1}
\end{table}
For sufficiently large $t$ (or $\tau$), one can define a probability
per unit proper time
\be
{\mathcal P}_{2\,\Omega}(R,\dot R)
\equiv
{P(2\,\Omega;\tau)\over \tau}
\simeq
e^2\,\dot R^2\,F(R,M_s)\,
\delta\left(2\,\Omega-\omega_{n0}/\sqrt{f_{\rm o}}\right)
\ ,
\label{P}
\ee
where the function $F$ can be straightforwardly deduced
from Eq.~(\ref{P2O}).
The above expression shows that emissions occur when some frequency
$\omega_{n\,0}/\sqrt{f_{\rm o}}$ (as measured in the shell frame)
of an AdS mode is close to the proper energy gap $2\,\Omega$, that
is
\be
n^{3/2}\sim {M\,M_s\over N\,\ell_p\,R}
\ .
\ee
\par
From the probability (\ref{P}) the energy emitted per unit time
is straightforwardly obtained as
\be
\dot M_s=-8\,\pi\,R^2\,\Omega\,{\mathcal P}_{2\,\Omega}(R,\dot R)
\ ,
\ee
which, together with the dynamical equation (\ref{Req}) constitutes
the system of coupled equations that we solved numerically.
In Fig.~\ref{R} we plot an example of the radius, velocity and ADM
mass of the shell for a choice of the parameters particularly suited
to obtain neat plots.
It is clear that the velocity approaches a terminal value, although
our approximations are not entirely reliable for too long times.
For more realistic values of the parameters, it suffices to estimate
some relevant quantities at $\tau=0$ that we give in Table~\ref{T1}.
It is clear from those figures that the larger $\ell$, the
smaller is the effect (at small times).
\par
To summarize, we have analyzed the semiclassical dynamics of a
shell made of bosons in AdS, thus generalizing the results previously
obtained in asymptotically flat space-time \cite{acvv2}.
As in the latter case, we find that the shell spontaneously emits
gravitational energy in the form of radiation when its velocity
of contraction is not negligible and the adiabatic approximation
fails.
In AdS this occurs not only for the shell radius $R$ approaching the
gravitational (Schwarzschild) radius $R_H\sim 2\,M_s$ but also for
relatively large radii, $R\gg\ell$, thus increasing the importance
of such an effect.
Another peculiarity of AdS is that the spectrum of massless radiation
is not continuous, and one therefore finds that the emission occurs
as a resonance between the shell inner energy gaps and the AdS
levels.
Our approach therefore suggests that the quantum nature of the
collapsing matter implies the existence of radiation and possibly
strong backreaction.
Hence, the choice of vacua in Ref.~\cite{giddings} for a shell radius
much larger than the horizon radius (Boulware vacuum) and for a shell
approacing the horizon (Unruh vacuum) may actually be connected by
the evolution of the system when the backreaction is properly
considered.
Since the radiation is present from the very initial stages of the
collapse, this situation is analogous to the ``topped-up'' Boulware
state used in Ref.~\cite{israel2} with a radiation of dynamical origin.
\end{document}